\documentclass[preprint2]{aastex}

\usepackage{natbib}

\shorttitle{Impact of uncertainties in stellar parameters on integrated spectra}
\shortauthors{Percival \& Salaris}

\begin{document}


\title{The impact of systematic uncertainties in stellar parameters on integrated spectra of stellar populations}


\author{Susan M. Percival and Maurizio Salaris}
\affil{Astrophysics Research Institute, Liverpool John Moores University, Twelve Quays House, Birkenhead, CH41 1LD}
\email{smp,ms@astro.livjm.ac.uk}


\begin{abstract}
In this paper we investigate a hitherto unexplored source of potentially 
significant error in stellar population synthesis (SPS) models, caused by 
systematic uncertainties associated with the three fundamental stellar
atmospheric parameters; effective temperature $T_{eff}$, surface gravity $g$, 
and iron abundance [Fe/H].  All SPS models rely on calibrations of 
$T_{eff}$, log$g$ and [Fe/H] scales, which are implicit in stellar models, 
isochrones and synthetic spectra, and are explicitly adopted for empirical 
spectral libraries.  We assess the effect of a mismatch in scales between 
isochrones and spectral libraries (the two key components of SPS 
models) and quantify the effects on 23 commonly used diagnostic line indices.
We find that typical systematic offsets  of 100K in $T_{eff}$, 0.15~dex in
[Fe/H] and/or 0.25~dex in log$g$ significantly alter inferred absolute ages 
of simple stellar populations (SSPs) and that in some circumstances,
relative ages also change.  Offsets in $T_{eff}$, log$g$ and [Fe/H] scales
for a scaled-solar SSP produce deviations from the model which can 
mimic the effects of altering abundance ratios to non-scaled-solar chemical 
compositions, and could also be spuriously interpreted as evidence for a
more complex population, especially when multiple-index or full-SED fitting 
methods are used.  We stress that the behavior we find can potentially affect 
any SPS models, whether using full integrated spectra or fitting functions
to determine line strengths.  We present measured offsets in 23 diagnostic 
line indices and urge caution in the over-interpretation of line-index data 
for stellar populations.
\end{abstract}


\keywords{stars: evolution --- galaxies: evolution --- galaxies: stellar content}



\section{Introduction}
\label{sec:intro}

In recent years, stellar population synthesis (SPS) models have become a 
fundamental tool in the study of both Galactic (resolved) and extra-galactic 
(unresolved) stellar populations.  Predictions of both photometric and 
spectroscopic properties,
in terms of broad-band colors and line index strengths, from various SPS 
models have yielded methods for tackling the well known age-metallicity
degeneracy in simple (single-age, single metallicity) stellar populations 
(SSPs -- see e.g. \citealt{basti4,galev09,coelho07,starb99,maraston05,bc03} 
for some recent examples).

As observational data, and in particular spectroscopic data, are becoming more
precise, the models are becoming more sophisticated and methods are being 
developed to determine more detailed information such as the level of 
$\alpha$-enhancement in SSPs, chemical evolution, and the general properties of 
the star formation histories of composite populations.  However, results
from different sets of models often disagree and there are currently many
unresolved issues concerning their interpretation and implementation.  
\citet{carter} investigated the ability of several ``off the shelf'' SPS 
models to reproduce optical and near infra-red colors (from $u$ through to $K$)
of a small sample of well-studied nearby elliptical galaxies.  Fitting to
SSPs, making the implicit assumption that giant ellipticals can be well 
represented by a simple population with no significant star 
formation history, they went on to compare the ages and metallicities 
predicted from the various models, using 
a simple reduced $\chi^{2}$ test.  Carter et al. found that, for a 
similar $\chi^{2}$, best-fits from different SPS models can give ages which 
differ by 5~Gyr or more whilst predicted [Fe/H] can easily differ by 0.4 dex 
(a problem here being the [Fe/H] sampling, which for most sets of SPS 
models is limited to $\sim$0.3 dex).
Predictions from spectroscopic data, in the form of diagnostic line indices, 
can have similar (if not worse) discrepancies, as demonstrated by the 
results of the
IAU Symposium 241 Stellar Population Challenge\footnote{http://www.astro.rug.nl/$\sim$sctrager/challenge/}.  
For the well-studied Galactic Globular Cluster 
47~Tuc, which was set as a test object with an observed spectrum provided
from \citet{schi05}, predicted ages from various 
models range from 5~Gyr to $>$18~Gyr whilst predicted [Fe/H] ranges from 
$-0.1$ to $-0.8$ dex.

Several recent studies have focussed on discrepancies that can arise because 
of properties of the observed population, which may not be fully accounted 
for in the models.  These include extended blue horizontal branches and blue 
stragglers which can both make a population look spuriously young 
(\citealt{schi04,cenarro}, respectively, using diagnostic line indices), and 
thermally pulsing AGB stars which strongly affect the near-infrared flux
(and hence colors) at intermediate ages, significantly altering the 
inferred galaxy masses \citep{bruzual}.
However it is clear that there can be significant offsets between ages
and metallicities determined from (or predicted by) different sets of SPS
models which must, at least in part, be due to systematic differences in the 
model ingredients. 
This problem is clearly demonstrated by the `zero-point' offset in some 
models utilizing diagnostic line indices
which means that inferred ages of Galactic globular clusters can be older than 
14~Gyr.  Although these problems have been noted by several groups 
\citep[see e.g.][]{vaz01,schi02,cenarro} little work 
has been done to determine the specific sources of these offsets or quantify 
their effects in terms of the true systematic errors which should be assigned 
to derived ages and metallicities. 

Physical parameters of stars within a population, such as effective 
temperature, $T_{eff}$, surface gravity, log$g$, and iron abundance, [Fe/H], 
are key parameters in the construction of the SPS models 
(see Section~\ref{sec:ingred}) and one, as yet, unexplored aspect of SPS is 
the adopted temperature and metallicity scales used in the underlying 
stellar models and spectra.  When generating stellar evolution models 
(which give rise to isochrones, a fundamental building block in SPS modelling),
the temperature and metallicity scales effectively come out of the models 
themselves.  This is also the case for synthetically generated stellar
spectra, for which stellar atmosphere models and adopted line lists give rise 
to some intrinsic temperature and metallicity scale.  For empirical spectra, 
a specific temperature and metallicity scale must be adopted in order to
determine physical parameters of the stars.  There can potentially be 
mismatches between the temperature and metallicity scales of the underlying
stellar models and those of the stellar spectra (whether synthetic or 
empirical), but any impact this may have 
on the resulting SPS models has not yet been investigated systematically and 
quantified.  The aim of this paper is to address this issue.  

The structure of the paper is as follows:  
In Section~\ref{sec:ingred} we review the ingredients required to construct 
SPS models whilst Section~\ref{sec:prelim} describes some preliminary tests 
which motivated this work. 
Sections~\ref{sec:tests} and \ref{sec:results} describe the specific tests 
applied to the models and the results of these tests whilst 
Section~\ref{sigerrors} briefly assesses the significance of the results in 
the context of typical observational errors for extragalactic systems.
In Section~\ref{sec:discsum} we summarize our results and discuss their 
implications for the age and metallicity predictions of SPS models and 
their associated errors.

\section{Model ingredients and construction}
\label{sec:ingred}

Our work focusses on the use of diagnostic line indices
to determine ages and metallicities for SSPs -- in particular we
will utilize the commonly adopted technique of comparing plots of pairs of 
line indices with model grids in our investigation, although our results also
have important implications for other methods, e.g. 
multiple-index fitting or whole-spectrum fitting.

Throughout this work we will be using the high resolution synthetic spectra
from the BaSTI SPS models\footnote{http://albione.oa-teramo.inaf.it/}, 
described in \citet{basti4}.  Since we 
have control over all the ingredients of these models, we can separately 
investigate the effects of each input element on the resulting spectral energy
distributions (SEDs) and line indices.

The two principal ingredients of any SPS model are the underlying stellar 
models, in the form of isochrones, which are `populated' according to some 
initial mass function (IMF)
to create each stellar population, and the spectral library which is used to 
assign a spectrum to each data point in that population.  The summing of these
individual spectra, with appropriate weighting, results in the final integrated
spectrum (SED) on which diagnostic line strengths, such as H$\beta$ and 
various Fe and other metal lines, can be measured directly as equivalent 
widths (EWs).  This was the method employed to
produce the BaSTI database of SEDs, and is used here to test the models 
\citep[see][for more details]{basti4}. 
Alternatively, using methods pioneered by \citet{worthey94}, some groups use
fitting functions to determine line strengths \citep[see e.g.][]{tmb03,schi07,trager08}.  
Our study is also relevant to these methods since 
they also rely on calibrations from spectral libraries and are constructed 
in a similar way, i.e. by assigning the relevant quantity (EW or similar) to
each point in a population and then summing along the isochrone.

The choice of isochrones used by SPS modellers is undoubtedly a contributory 
factor to the differing predictions from various groups.  Underlying stellar 
models (and hence isochrones) from different sources can vary because of 
the specifics of assumed input physics and differences in stellar 
evolution codes.  An in-depth discussion of these differences is beyond the 
scope of this paper, however a detailed comparison of some of the most commonly 
used isochrone sets can be found in \citet{basti1}.  One practical difference
between isochrones from different groups is the number of evolutionary 
points (EPs) that define each isochrone -- effectively the sampling along the
isochrone.  BaSTI isochrones consist of 2250 discrete EPs covering all 
evolutionary stages whilst, for example, the commonly used Padova isochrones
are typically defined by a few hundred EPs 
\citep[e.g. those of][]{girardi2000}.  

Evolutionary points along isochrones are defined in terms of effective 
temperature $T_{eff}$, luminosity $L$, and mass $M$, from which surface 
gravity, log$g$, is also derived.  
$T_{eff}$ and log$g$, along with [Fe/H] (and/or total $Z$ and 
degree of $\alpha$-enhancement) are used to match, or create by interpolation, 
an appropriate spectrum (or fitting function) to each EP, since spectra in 
libraries are usually parametrized by these 3 quantities.  
$T_{eff}$ and [Fe/H] (and by implication log$g$)
must be defined on some scale, and there can potentially be mismatches between 
the scales arising from the underlying stellar models/isochrones and those 
adopted by the spectral library used. This is especially relevant when 
empirical 
spectral libraries are employed as this requires an evaluation of the
physical parameters of the observed stars, however it is also an issue when 
using synthetic spectral libraries if the stellar models and the atmosphere 
models are on different scales.  Any mismatches in these scales will 
equally affect SPS models using either full SEDs or fitting functions.

\section{Preliminary tests}
\label{sec:prelim}

As a preliminary evaluation of the typical systematic offsets to be 
expected in effective 
temperature scales, we compared the listed $T_{eff}$ values for EPs along 
BaSTI isochrones with those that would be `predicted' from the empirical 
$T_{eff}$--color calibrations of \citet{alonso96} (for dwarfs and subdwarfs) 
and \citet{alonso99} (for giants).  The \citet{alonso96,alonso99} 
calibrations give equations to calculate $T_{eff}$ from various colors, in 
combination with [Fe/H] -- for this test the $(V-K)$ color calibration
was used, since this is the one upon which the atmospheric parameters for 
the MILES empirical spectral library \citep[][used in a later test]{miles1} are
based.  Taking the $(V-K)$
color for each EP along an isochrone \citep[see][for details of isochrone 
colors]{basti1}, along with [Fe/H], $T_{eff}$s were calculated from the Alonso
et al. calibrations and then compared with the listed isochrone $T_{eff}$s.
Systematic differences were found, which have some metallicity dependence, 
and are typically around 70K at solar metallicity and up to 160K at 
${\rm [Fe/H]}=-2$, in the sense that the listed isochrone $T_{eff}$s are 
hotter.  These differences are similar to the typical quoted systematic
uncertainties on $T_{eff}$, which are of the order 50--100K 
\citep[see e.g.][]{alonso96}. 

\citet{ramirez} determined temperatures of 135 dwarf and 36 giant FGK stars
using a method very similar to the Alonso et al. studies and found that, in
general, the agreement in temperature scales was very good, and the mean 
uncertainty in derived temperatures is 75K for dwarfs and 60K for giants. 
However, \citet{casagrande} found that for 18 stars in common between their 
study and that of \citet{ramirez}, there was an average difference of 
105$\pm$72K, with the \citet{casagrande} scale being hotter.  
\citet{casagrande} go on to say `though not negligible, such differences are 
within the error bars of current temperature determinations'. 
More recently \citet{gonz} rederived $T_{eff}$ for all stars in the Alonso et 
al. samples \citep{alonso96b,alonso99b} using the infrared flux method, and 
found differences in temperature scales of $\sim$64K and $\sim$54K for 
dwarfs and giants respectively, in the sense that the \citet{gonz} scale 
is hotter.  However, comparing with the \citet{ramirez} calibration they 
found that for low metallicity dwarfs, the \citet{gonz} scale is cooler by
$\sim$87K.

[Fe/H] and log$g$ calibrations are, 
of course, intrinsically linked to $T_{eff}$ when determining atmospheric 
parameters for empirical spectra and so systematic uncertainties in 
$T_{eff}$ imply similar uncertainties on these parameters also.  Quoted
uncertainties are typically of the order 0.25 dex in log$g$ and 0.15 dex in 
[Fe/H] \citep[see e.g.][]{skc}.  

In the early stages of creating integrated spectra for the BaSTI SPS 
database we performed a preliminary comparison between synthetic spectra 
from the \citet{munari} spectral library (as used for the high resolution 
BaSTI SEDs)and the MILES empirical spectral library 
\citep[][as used by \citealt{cenarro}]{miles1}.  For this exercise, we
took a subset of the MILES spectra with zero reddening ($E(B-V)=0.0$) and
with the designation SKC, indicating that the atmospheric parameters
are from \citet{skc} (these stars being the `gold standard' for the MILES 
library).  To avoid the ambiguity of whether or not to assume scaled-solar 
or $\alpha$-enhanced synthetic spectra, we used only stars with either 
[Fe/H]$\geq -0.3$ (assumed to be scaled solar) or [Fe/H]$\leq -1.0$ 
(assumed to be $\alpha$-enhanced -- note that in the Munari library, the
level of $\alpha$-enhancement is fixed at [$\alpha$/Fe]=0.4). 
By interpolating in 
[Fe/H], $T_{eff}$ and log$g$ amongst spectra in the \citet{munari} library, 
a matching synthetic spectrum was created for each star in the MILES 
subsample.  Line strengths were then measured directly on the two sets of 
spectra using the LECTOR program of 
A. Vazdekis\footnote{see http://www.iac.es/galeria/vazdekis/} (as done in
\citealt{basti4}) and a mean offset found of $\sim 0.4$~dex in H$\beta$, the
principal age indicator for SSPs, in the sense that the MILES spectra have
larger values.  Significant mean offsets were also found in the main 
metallicity indicators, e.g. $\Delta{\rm Mgb} \simeq 0.6$,  
$\Delta{\rm [MgFe]} \simeq 0.5$ and 
$\Delta{\rm Fe5406} \simeq 0.2$, in the sense that the MILES
spectra have smaller values.  One possible interpretation is that these
systematic offsets are related to the calibration of atmospheric 
parameters, which may differ between the two spectral libraries.
The systematic offsets that we find are qualitatively similar to those found
by \citet{martco}, who performed a comparison of spectral indices measured
on several theoretical and empirical spectral libraries, using a slightly
different method to ours.  \citet{martco} generally found that disagreements 
between spectral libraries have some temperature dependence, and that the
largest offsets often occur for the cooler stars (defined as $T_{eff}<4500$K 
in their work).  
Exploring the impact that these differences can have on SPS models is a 
key motivation for the work presented in this paper.

In the following tests we examine the effects of altering each of the
atmospheric parameters in the the BaSTI SPS models separately, within the 
typical uncertainties, whilst holding all other elements constant -- these 
uncertainties are taken to be $\pm$100K in $T_{eff}$, $\pm$0.25 dex in log$g$ 
and $\pm$0.15 dex in [Fe/H].  
We stress that these tests are exploring the effects of a {\it mismatch} 
in scales between the underlying stellar models (and isochrones) and the 
adopted spectral library in the SPS models, which potentially give rise to
systematic errors in derived ages and metallicities for SSPs.
When presenting our results we are making the implicit assumption that
the differential behavior we find would be quantitatively the same, or
very similar, for other SPS models, whether based on full integrated spectra
or on fitting functions methods to determine line strengths.

\section{The tests}
\label{sec:tests}

All the tests described below were performed on two test-case SSPs with ages
$t$=14~Gyr, to represent a typical old elliptical galaxy, and $t$=4~Gyr,
which is representative of the intermediate ages found in the sample
of  ellipticals studied by \citet{trager2000}.
In both cases the scaled-solar abundance, solar 
metallicity models were used ($Z=0.0198$ and ${\rm [Fe/H]}=+0.06$) with no
convective overshooting, and with the Reimers mass-loss parameter, $\eta$, set
at 0.2.  High resolution (1{\rm \AA}/pixel) integrated spectra for the 
reference models were 
created, as described in Percival et al. (2009), where the interested reader 
can also find more details on the underlying stellar models.  We note here 
that we have also applied all the following tests to low metallicity 
$\alpha$-enhanced SSPs, typical of Galactic globular clusters, and obtain
quantitatively very similar results.

When constructing the integrated spectrum for each test, the underlying 
isochrone is left unchanged, but individual spectra are assigned to each 
EP along the isochrone with the appropriate offset in each atmospheric 
parameter, as listed below.  
Each atmospheric parameter was tested for two instances -- one in which the
whole isochrone is affected (i.e. stars at all evolutionary phases), and one 
in which only the giant stars are affected.  For these purposes we took all 
EPs with, simultaneously, $T_{eff}<$5000K and log$g<$3 to represent giants.
The justification for treating giants only as a separate case is that,
for empirical $T_{eff}$ calibrations, dwarfs and giants are often treated
separately and consequently have separate $T_{eff}$--color relationships.
%\citep[see][respectively]{alonso96,alonso99}.  
It is also generally the case that cool stars are harder to model 
theoretically due to the effects of phenomena such as molecular opacities,
convection and mass loss which become increasingly important for cooler 
stars, hence there is more 
likely to be a discrepancy in temperature scales for cool RGB and AGB stars.  
Tests were performed on the two reference SSPs as follows:

Test 1 -- Spectra assigned with $T_{eff}$ increased by 100K compared to 
isochrone $T_{eff}$s (whole isochrone).

Test 2 -- Spectra assigned with $T_{eff}$ increased by 100K compared to 
isochrone $T_{eff}$s, for giants only (i.e. $T_{eff}<$5000K and log$g<$3 on 
isochrone).

Note that we could not perform fully equivalent tests in which $T_{eff}$ was 
decreased by 100K.  
This is because there is a low temperature cut off in the spectral 
library used for the BaSTI SPS models \citep[][see \citealt{basti4} for more 
details]{munari} which means that the coolest stars could not be consistently 
included in the test as an offset of --100K would take them below this 
threshold.
However, we did perform this test using a truncated version of the 4~Gyr SSP,
i.e. with a cut at 100K above the threshold $T_{eff}$, so that this test
could be done by assigning spectra with both increased and decreased  
$T_{eff}$ values.  

Tests 3a,b -- log$g$ increased/decreased by 0.25 dex (whole isochrone)

Tests 4a,b -- log$g$ increased/decreased by 0.25 dex (giants only)

Tests 5a,b -- [Fe/H] increased/decreased by 0.15 dex (whole isochrone)

Tests 6a,b -- [Fe/H] increased/decreased by 0.15 dex (giants only)

Two further tests were also performed in which two `worst-case' scenarios 
were created.  For the first of these, two parameters were altered 
simultaneously over the whole isochrone, namely $T_{eff}$ increased by 100K 
and log$g$ decreased by 0.25 dex.  For the second of these tests all three
parameters were altered -- $T_{eff}$ was increased, log$g$ decreased and [Fe/H]
increased by 0.15~dex.  The purpose of these tests was to check 
whether the effects on line indices are additive when two or three atmospheric 
parameters are offset simultaneously.

Line strengths for 23 diagnostic indices were then measured for the reference 
cases and test cases in the same way as before, i.e. by measuring 
equivalent widths directly on the integrated spectra themselves.  
Indices used are the 21 indices defined by the bandpasses in 
\citet{trager98} (noting that, by convention, CN$_{1}$, CN$_{2}$, Mg$_{1}$,
Mg$_{2}$, TiO$_{1}$ and TiO$_{2}$ are quoted in magnitudes rather
than as EWs in {\rm \AA}ngstroms) plus H$\delta_F$ and H$\gamma_F$ as defined
in \citet{worthott}.  We stress that all numbers quoted are line strengths
as measured directly on the integrated spectra and are {\it not} transformed 
onto the Lick system. 

\section{Results}
\label{sec:results}

The results of the tests on the 4~Gyr and 14~Gyr populations are tabulated 
in Tables~\ref{tab:4Gtests} and \ref{tab:14Gtests} respectively, where the 
second column of each table shows the measured index strengths for the 
unaltered reference SSP, and subsequent columns list the offset in each index 
between the reference SSP and each of the spectra resulting from tests 1--6. 
The quantities listed are offsets which must be {\it added} to the 
reference values to reproduce the results of each test.  

\subsection{Effects of altering $T_{eff}$}
\label{subsec:teff}

Results from tests 1 and 2, i.e. $T_{eff}$ increased for the whole isochrone 
and for giants only, respectively, are shown in the third and fourth columns of
Tables~\ref{tab:4Gtests} and \ref{tab:14Gtests}.  
For both test-case SSPs it can be seen that H$\beta$ increases significantly 
when $T_{eff}$ is increased for the whole isochrone, as would be expected.  
However when $T_{eff}$ is increased for giants only, the H$\beta$ value 
increases slightly  for the 14~Gyr SSP and actually {\it decreases}
for the 4~Gyr SSP -- we attribute this to the fact that for giant stars at 
these low temperatures ($T_{eff}<5000$K) and gravities (log$g<$3), increasing 
$T_{eff}$ by 100K increases the continuum level around the H$\beta$ line 
but does not significantly increase the strength of the line itself, and 
so the measured H$\beta$ value can decrease.  The higher order Balmer 
lines H$\delta_F$ and 
H$\gamma_F$ behave in a qualitatively similar way to H$\beta$ in that they
increase significantly when $T_{eff}$ is increased for the whole isochrone,
but increase only marginally, or decrease slightly, for the giants-only case.
 
For all but one of the metal indicators (including Mg$b$ and all the Fe lines)
index strengths decrease when $T_{eff}$ is increased for the whole isochrone,
the exception being C$_2$4668 for the 4~Gyr SSP, which increases slightly.  
When $T_{eff}$ is increased for giants only, offsets are generally smaller,
with the exception of C$_2$4668, but for many indices offsets go in the 
opposite sense to those for the whole-isochrone case, e.g. CN1, CN2 and 
most of the Fe lines. 

Results of the decreased $T_{eff}$ test (using the truncated 4~Gyr SSP) are 
not listed, as the measured values of offsets would not be applicable to the
full SSP.  However we note that, for all 23 measured indices, offsets for the 
decreased $T_{eff}$ truncated SSP were found to be symmetrically opposite 
with respect to the increased $T_{eff}$ truncated SSP case, 
implying that this would also be the case for the full SSP.  

\begin{figure}
\epsscale{1.00}
\plotone{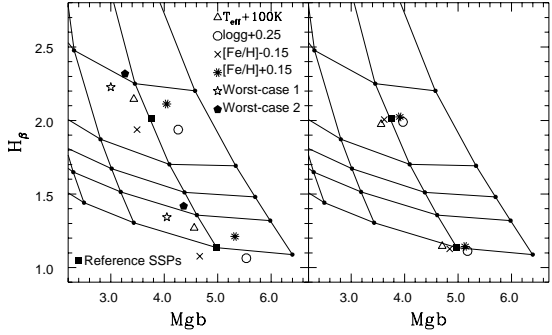}
\caption{H$\beta$ vs. Mg$b$ grid for BaSTI scaled-solar SSP models for ages 
3, 6, 8, 10 and 14~Gyr (age increasing from the top downwards) and 
[Fe/H]$=$ $-$0.66, $-$0.35, +0.06 (solar) and +0.40 (increasing from left to 
right). The left-hand panel shows results of the whole-isochrone tests
and the right-hand panel shows results of the giants-only tests.
The two test reference SSPs (solar metallicity, 4~Gyr and 14~Gyr) are 
marked as solid squares.  Results of various tests are labelled on the 
diagrams (see text for details).}
\label{fig1}
\end{figure}

\begin{figure}
\epsscale{1.00}
\plotone{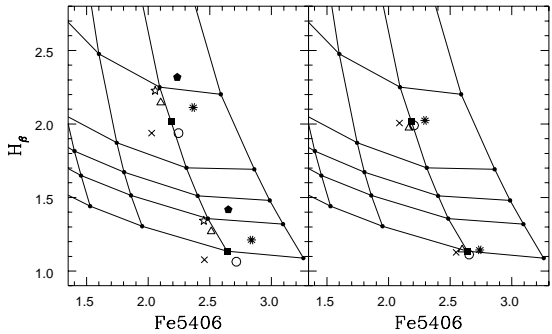}
\caption{H$\beta$ vs. Fe5406.  Panels, grid points and symbols are the same 
as for Figure~\ref{fig1}.}
\label{fig2}
\end{figure}

\begin{figure}
\epsscale{1.00}
\plotone{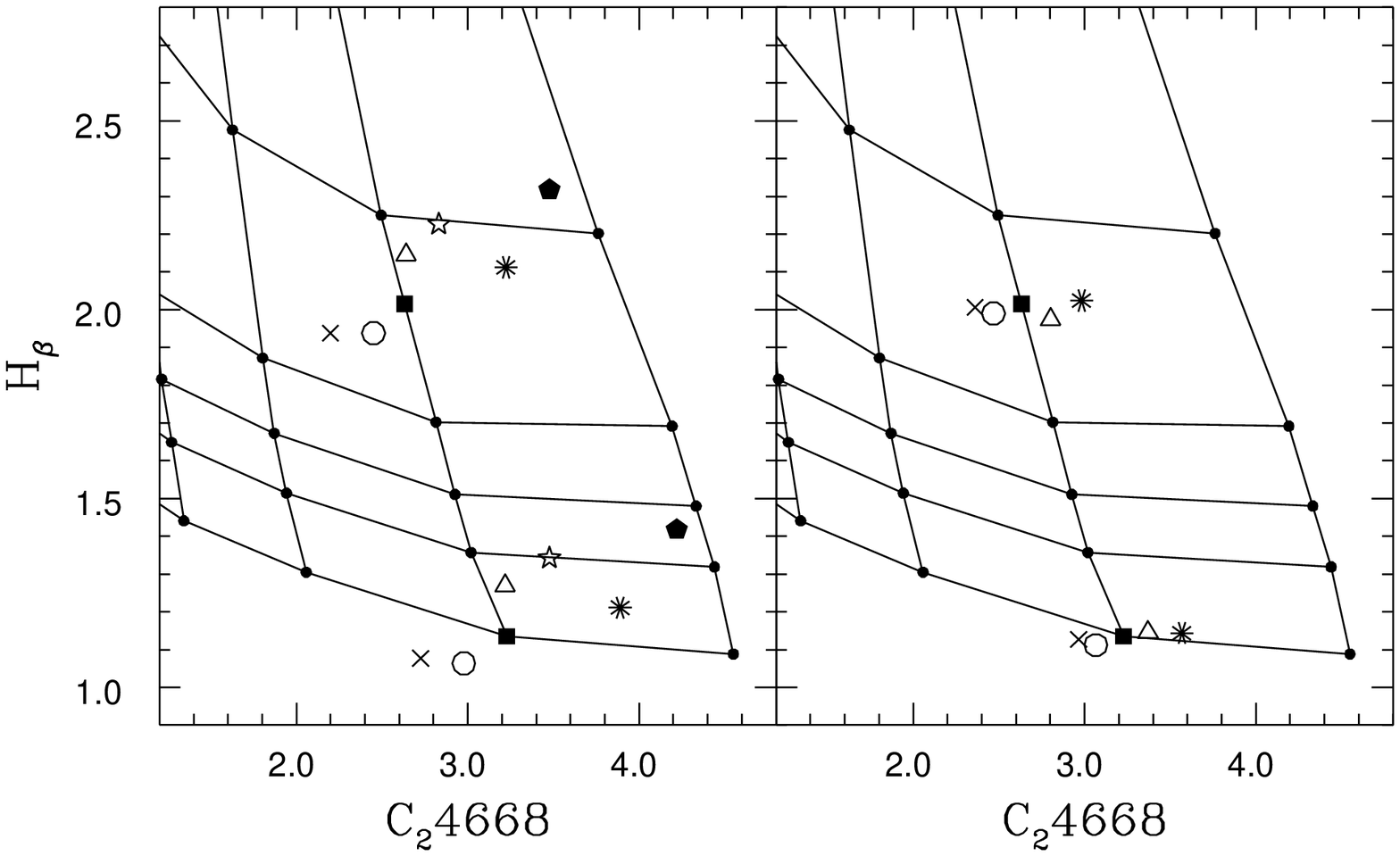}
\caption{H$\beta$ vs. C$_2$4668. Panels, grid points and symbols are the same
as for Figure~\ref{fig1}.}
\label{fig3}
\end{figure}

To demonstrate the effect of these offsets on estimated ages and metallicities
of SSPs, we have constructed representative index-index grids from the BaSTI 
models for a range of ages and metallicities and overplotted the test-case 
SSPs -- Figures~\ref{fig1}, \ref{fig2} and \ref{fig3} show grids
of H$\beta$ vs. Mg$b$, Fe5406 and C$_2$4668, respectively.  In each figure
the left hand panel shows the results of the whole-isochrone tests and the
right hand panel shows the results of the giants-only tests.  It can be seen
that increasing $T_{eff}$ by 100K for the whole isochrone makes the
apparent age of the 4~Gyr SSP approximately 0.75~Gyr younger whilst the 
14~Gyr SSP looks 2.5$-$3~Gyr younger, an apparent reduction in age of 
$\sim$20\% for both populations.  The effect on apparent ages for the 
giants-only test is small, altering absolute ages by no more than $\sim$3\%, 
however since the offsets for the 4~Gyr and 14~Gyr populations go in 
opposite senses, the relative ages also change.

Assessing the effect of temperature changes on derived metallicities is more 
complicated, since it depends on which index is used and whether metallicity
is defined in terms of [Fe/H] or total metallicity, $Z$.  The three grids
displayed in Figures~\ref{fig1}, \ref{fig2} and \ref{fig3} demonstrate the
problem.  

\begin{figure}
\epsscale{1.00}
\plotone{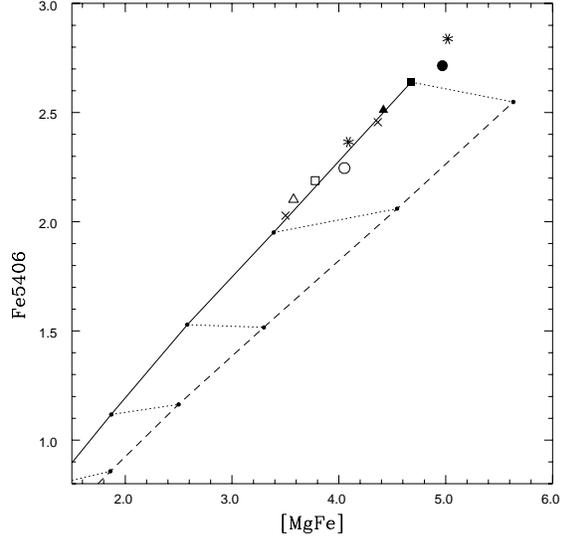}
\caption{Fe5406 vs. [MgFe] diagram for 14~Gyr SSP scaled solar (solid line) 
and $\alpha$-enhanced (dashed line) models, joined at approximately equal 
[Fe/H] (dotted lines) for [Fe/H] $\sim +0.06, -0.3, -0.7, -1.0$ (decreasing
from the top downwards).  Test 
reference SSPs are marked as squares, and other symbols are as for 
Figure~\ref{fig1}, for the 4~Gyr (open symbols) and 14~Gyr (solid symbols)
cases. }
\label{fig4}
\end{figure}
For a 100K increase in $T_{eff}$, the inferred [Fe/H] from the 
H$\beta$--Mg$b$ plane is $\sim$~0.1~dex lower than the reference SSP,
from the H$\beta$--Fe5406 plane it is virtually unchanged, and from the
H$\beta$--C$_2$4668 plane [Fe/H] apparently increases by $\sim$~0.05 dex.  
Offsets are similar for both the `whole-isochrone' and `giants-only' cases.  
The H$\beta$--C$_2$4668 grid also demonstrates another important point -- 
for the 14~Gyr SSP, the offset for the increased 100K case {\it decreases}
the absolute value of C$_2$4668 slightly, however the inferred [Fe/H] 
{\it increases} as this grid is not orthogonal (as is also the case for all
other commonly used age/metallicity diagnostic grids).  Similarly constructed 
grids (not displayed here) show that, with increasing $T_{eff}$, Ca4227 
behaves in the same way as Mg$b$, [MgFe] behaves as Fe5406 (but traces total
$Z$ rather than [Fe/H] -- see \citealt{basti4}) and CN1 and CN2 behave 
similarly to C$_{2}$4668.  The issue of the
contradictory behavior of various diagnostic metal lines and the impact on
inferred chemical composition will be discussed further in 
Section~\ref{sec:discsum}.

\subsection{Effects of altering log$g$}
\label{subsec:logg}

It was found that, for the log$g$ tests, offsets in all indices were found 
to be symmetrical for the increased/decreased cases with respect to the 
reference case, hence the results are listed in Tables~\ref{tab:4Gtests} and 
\ref{tab:14Gtests} as single values for each pairing of tests 3a/3b and 4a/4b, 
with the appropriate $\pm$ or $\mp$ sign.  For both test-case SSPs it can be 
seen that H$\beta$, H$\delta_F$ and H$\gamma_F$ increase significantly when 
log$g$ is decreased by 0.25~dex (and decrease when log$g$ is increased), 
affecting derived ages in a quantitatively similar way to a 100K increase in 
$T_{eff}$.  
Mg$b$, Ca4227, Fe5406, [MgFe] and C$_2$4668 also behave in a 
quantitatively very similar way with decreasing log$g$ to 
the increased $T_{eff}$ case.  Figures~\ref{fig1}, 
\ref{fig2} and \ref{fig3} show that, with increasing log$g$, [Fe/H]
inferred from the H$\beta$--Mg$b$ plane increases by $\sim$~0.1~dex, whilst 
that from the H$\beta$--[Fe5406] plane is virtually unchanged and from the
H$\beta$--C$_2$4668 plane [Fe/H] apparently decreases by $\sim$~0.1~dex.

\subsection{Effects of altering [Fe/H]}
\label{subsec:feh}

As for the log$g$ and $T_{eff}$ tests, the [Fe/H] tests produced offsets in 
all indices that were found to be symmetrical for the increased/decreased 
cases with respect to the reference case.  Results are listed in 
Tables~\ref{tab:4Gtests} and \ref{tab:14Gtests} as single values for each 
pairing of tests 5a/5b and 6a/6b, with the appropriate $\pm$ or $\mp$ sign.
As would be expected, all metal lines increase in strength when [Fe/H] is 
increased and decrease when [Fe/H] is decreased.  For the whole-isochrone 
tests, the inferred change in [Fe/H] from the diagnostic diagrams is 
consistent with the modelled change in [Fe/H], i.e. $\sim$~0.15~dex, 
whilst for the giants-only test the inferred change in [Fe/H] is generally 
about half this level.

An important point to notice is that H$\beta$ increases significantly when
[Fe/H] is increased over the whole isochrone (and decreases when [Fe/H] is
decreased) inducing $\sim$~12\% change in derived ages for a 0.15~dex change
in [Fe/H], however for the giants-only case H$\beta$ is virtually unchanged.
Rather puzzlingly, H$\delta_F$ and H$\gamma_F$ behave in the
opposite sense to H$\beta$ in that they both {\it decrease} significantly 
with increased [Fe/H], and vice versa. H$\delta_F$ and H$\gamma_F$ also 
display quantitatively similar offsets for the whole-isochrone and 
giants-only cases, unlike the situation for H$\beta$.
The implications of this contradictory behavior 
will be discussed further in Section~\ref{sec:discsum}.

\subsection{`Worst-case' scenarios}
\label{subsec:wcs}

The results from the `worst-case tests', in which two or three parameters 
were altered simultaneously, are not listed -- however in both cases the 
offsets in all indices were found to be additive. For the two parameter test 
(altering $T_{eff}$ and log$g$ simultaneously) resultant offsets are simply 
the sum of the offsets found for tests 1 and 3b, whilst for the three 
parameter test ($T_{eff}$, log$g$ and [Fe/H]) offsets are the sum of those from
tests 1, 3b and 5a.  Results of the two `worst-case' tests are also 
illustrated in Figures~\ref{fig1}, \ref{fig2} and \ref{fig3}.

% figures are in macro /userpc25/smp/HRS/Pind_plot_pap2, figs 12, 15, and 18
% which also includes CN1, CN2, Ca4227, Fe5270, Fe5335 
% and could construct [MgFe] if required.
% Fe5406/[MgFe] plot is in f5406_mgfe_14G_pap2plot
% which also includes <Fe>/[MgFe].

\section{Significance of results: comparison with typical observational errors}
\label{sigerrors}

In order to assess the significance of our results for observed stellar
populations, we briefly compare the line index offsets detailed above with 
typical observational errors for extragalactic systems.
Table~\ref{obserrors} shows the mean observational errors on various
line index strengths from three sources.
\citet{sanbla06} present a study of 98 early type 
(E and S0) galaxies in the local field, small groups, and some Virgo and 
Coma cluster members, whilst \citet{trager08} study 12 elliptical and S0 
galaxies in the Coma cluster.  For higher redshift systems we used data 
from \citet{sanbla09}, which comprises a catalogue of 215 red sequence 
galaxies in clusters and groups, with redshifts between 
$z \sim 0.45$ and $z \sim 0.75$.  For the \citet{sanbla09} sample, 
observational errors on all quoted indices are slightly larger than the 
largest offsets found in our work, so that the systematic effects that we 
find would be a significant, but not the dominant, source of error for these 
systems. However, for the more local systems observational errors for 
all the key diagnostic indices are significantly less than the offsets 
caused by the mismatches in stellar parameters detailed above.  Most
importantly, the observational errors on H$\delta_{F}$,  H$\gamma_{F}$ and
H$\beta$ are smaller by a factor of $\sim 2$ than the offset induced by a
100K mismatch on temperature scales, which would lead to a systematic shift
in the inferred ages for these systems.  Similarly, observational errors on
Ca4227, Mg$b$ and most of the Fe lines are significantly smaller than the
offsets in indices caused by a mismatch in log$g$ or [Fe/H] scales.
It should be borne in mind that any offsets resulting from stellar parameter 
mismatches induce systematic rather than random errors.

We remind the reader that all our results discussed so far pertain to
intermediate age and old stellar populations.  At the suggestion of the 
referee we also performed several tests on a 500~Myr, solar metallicity 
population.  At this age (and younger) stellar populations do not
have a red giant branch as such, and so we performed the basic tests
of increasing $T_{eff}$ by 100K, decreasing log$g$ by 0.25~dex and altering
[Fe/H] by 0.15~dex, for the whole isochrone case only.  The offsets in indices
found for this younger age population are generally a factor $\sim 2$ 
smaller than those found for the 4~Gyr and 14~Gyr SSPs and so in most cases 
are less significant than the observational errors quoted in 
Table~\ref{obserrors}.  Also, the strength of the Balmer lines increases
much more rapidly with decreasing age for ages below $\sim 1$~Gyr and so 
even a substantial change in H$\beta$ of 0.2~dex (larger than any of the 
offsets found here) only changes the inferred age by $\sim 10-20$~Myr.

\subsection{Effect on broad-band colors}
\label{sec:cols}

The aim of this paper is primarily to investigate the effect of mismatches
in stellar parameters on diagnostic line indices, however we have also
made a brief study of the effect on broad-band colors for the main tests
outlined in Section~\ref{sec:tests}.  In general, a mismatch
in the log$g$ scale has a negligible effect on colors whilst the effects
of a 0.15~dex mismatch in [Fe/H] are small and are typically within the 
likely photometric errors, i.e. $\lesssim 0.02$ mag in all the optical and 
near-infrared bands.  As might be expected, the effect of increasing 
$T_{eff}$ by 100K (Test 1 above) is to make all the broad-band colors bluer.  
For the intermediate age and old populations studied here, the effect is 
small in the the optical bands, $\sim0.01$~mag in $(U-B)$ 
and $(B-V)$, and 0.02~mag in $(R-I)$, implying a minimal change in the 
inferred age, however the effect becomes greater at longer wavelengths, 
making the $(J-K)$ color bluer by $\sim0.04$ mag.  Hence broader baseline 
colors such as $(V-K)$ and $(B-K)$ are affected more strongly by an
increase in $T_{eff}$, the colors becoming bluer by at least 0.1~mag.  
As a result, a 100K increase in 
$T_{eff}$ can significantly decrease the inferred metallicity, rather than 
making a population look younger, if a color-color diagram such as 
$(V-I)/(V-K)$ or $(B-K)/(J-K)$  is used (see Figure 1 of, respectively,
\citealt{salcas} and \citealt{paj}).

\section{Summary and discussion}
\label{sec:discsum}

We have investigated a potentially significant source of systematic error
which can affect the ages and metallicities of stellar populations derived 
from SPS modelling, caused by systematic uncertainties in the three
principal stellar atmospheric parameters $T_{eff}$, log$g$ and [Fe/H].  
In practice, we have tested the effects of a mismatch in temperature, 
metallicity and gravity scales between the underlying isochrones and the
spectra (or fitting functions) used to construct SPS models.  We have
done this by constructing integrated spectra for two solar metallicity SSP 
models aged 4~Gyr and 14~Gyr, and applying offsets of 100K in $T_{eff}$, 
0.25~dex in log$g$ and 0.15~dex in [Fe/H], which we take as typical 
zero-point uncertainties in these key atmospheric parameters.  
We have then quantified the effect on various diagnostic line indices by 
measuring EWs on the resultant spectra and calculating offsets 
between the unaltered reference SSPs and the altered versions.

We note here that the magnitude of the offsets we find as a result of 
altering the stellar parameters for SSPs, as described, is quantitatively 
very similar to the differences in line index strengths found between the 
various spectral libraries, detailed in Section~\ref{sec:prelim}.  
Hence our results are consistent with the differences between the spectral 
libraries being largely caused by zero-point differences in their 
$T_{eff}$, log$g$ or [Fe/H] scales.  We take this as an indication that 
our results give a good estimate of the likely systematic errors in line 
index strengths inherent in any population synthesis model, due to possible 
zero-point mismatches in stellar atmospheric parameter scales.

Using simple index-index diagrams to make a preliminary assessment of the 
impact on stellar population parameters inferred from the models, we find 
that absolute ages derived from the H$\beta$ index can easily be affected at 
the 20\% level for both old and intermediate age populations.  Relative 
ages can also be affected, albeit at a lower level.  Inferred ages are more 
complicated to interpret if the $H\delta_F$ and H$\gamma_F$ indices are also
considered because of their behavior in response
to a systematic shift in [Fe/H], which goes in the opposite sense to that of 
H$\beta$.  The inferred systematic 
errors in [Fe/H] and/or $Z$ are also hard to quantify because of the 
opposite behavior of certain key metal indicators in response to systematic 
offsets in $T_{eff}$ and log$g$.  

This behavior has implications for methods which fit simultaneously to 
several indices (or perform a full SED fit) to derive ages and metallicities 
of stellar populations, since a failure to fit several indices simultaneously 
could, spuriously, be interpreted as an indication of non-solar abundance 
ratios.  Also, mismatches between observational data and model SSPs are often
taken as evidence for the presence of components which are not fully accounted 
for in the models, such as missing (or extreme) stellar evolutionary stages 
or a composite population.

\citet{basti4} demonstrated that, for SSPs, the Fe5406 index traces Fe only, 
whilst [MgFe] traces total metallicity, $Z$ (as first noted by 
\citealt{tmb03}).  These two indices, in combination therefore provide an 
estimate of the level of $\alpha$-enhancement, or simply whether a population 
has non-solar abundance ratios.  Figure~\ref{fig4} shows the Fe5406--[MgFe] 
plane with lines from the BaSTI 14~Gyr (constant age) scaled-solar and 
$\alpha$-enhanced models, joined at points of approximately equal
[Fe/H] (note that at constant $Z$, [Fe/H] is
lower for $\alpha$-enhanced models than the corresponding scaled-solar ones).
Overplotted are the results from the whole-isochrone tests for increased 
$T_{eff}$ (100K), increased log$g$ (0.25~dex) and increased/decreased [Fe/H] 
(0.15~dex).  In this diagram one expects that any deviation in abundance 
ratios will move points horizontally, i.e. any degree of $\alpha$-enhancement 
moves points from the scaled-solar line on the left, towards the 
$\alpha$-enhanced line on the right (see Figure 9 of \citealt{basti4} -- note
that lines of different age are completely degenerate in this diagram).  
Here it can be seen that altering any of the atmospheric parameters simply 
moves the scaled-solar SSP points along the scaled-solar line, changing the 
inferred [Fe/H] but not altering the inferred abundance ratios for other 
elements to non-solar ratios.  This is an important point to notice since 
individual element abundances, including several $\alpha$ elements, appear 
to alter substantially, as demonstrated by the measured offsets in the 
various line indices listed in Tables~\ref{tab:4Gtests} and \ref{tab:14Gtests}.

In conclusion, we urge caution against the over-interpretation of stellar
population parameters from line index data, in terms of the inferred 
scaled-solar or non-scaled-solar abundance ratios and also the inferred 
presence of a composite population, especially when multiple-index 
or full-SED fitting methods are employed.  We find that, for SSPs, 
Fe5406 in combination with [MgFe] provides the most robust indication of 
non-solar abundance ratios.  We remind the reader that our results 
potentially impact on all SPS methods,
whether fitting functions or full SEDs are employed.  Measured offsets for 
23 commonly used diagnostic line indices are provided, and we encourage 
the user to determine the overall impact on their observational data and
preferred fitting method.

\acknowledgments

We thank the anonymous referee for a constructive report and some useful
suggestions which helped to put our results in context.  
S.M.P. would like to express heartfelt thanks to Elaine 
Smith-Freeman for many useful discussions and for providing the initial 
motivation to do this work.
S.M.P. acknowledges financial support from the Science \& Technology 
Facilities Council (STFC) through a Postdoctoral Research Fellowship.

\begin{deluxetable}{lrrrrrrr}
\tablewidth{0pt}
\tablecaption{Results of tests on the 4Gyr SSP. The second column lists
EWs measured on the reference SSP spectrum and subsequent columns list the 
offset in each index between the reference spectrum and each of the test 
spectra (see text for details).  \label{tab:4Gtests}}
\tablehead{
\colhead{} & \colhead{} & \multicolumn{2}{c}{$T_{eff}$+100K} & \multicolumn{2}{c}{log$g \pm$0.25} & \multicolumn{2}{c}{[Fe/H]$\pm$0.15}\\
\colhead{} & \colhead{} & \multicolumn{1}{c}{Full iso} & \multicolumn{1}{c}{Giants} & \multicolumn{1}{c}{Full iso} & \multicolumn{1}{c}{Giants} & \multicolumn{1}{c}{Full iso} & \multicolumn{1}{c}{Giants}\\
\colhead{Index} & \multicolumn{1}{c}{Ref} & \multicolumn{1}{c}{(1)}  & \multicolumn{1}{c}{(2)} &
\multicolumn{1}{c}{(3a,b)} & \multicolumn{1}{c}{(4a,b)}  & \multicolumn{1}{c}{(5a,b)} & \multicolumn{1}{c}{(6a,b)}}
\startdata
H$\delta_{F}$ &    0.083  &  0.228   &  $-$0.109 &   $\mp$0.098 &   $\mp$0.011 &   $\mp$0.067 &   $\mp$0.081\\
H$\gamma_{F}$ & $-$1.208  &  0.315   &  $-$0.085 &   $\mp$0.181 &   $\mp$0.031 &   $\mp$0.078 &   $\mp$0.079\\
CN$_{1}$      &    0.002  & $-$0.003 &   0.007   &   $\mp$0.001 &   $\mp$0.001 &   $\pm$0.007 &   $\pm$0.005\\
CN$_{2}$      &    0.058  & $-$0.004 &   0.005   &   $\pm$0.000 &   $\pm$0.000 &   $\pm$0.009 &   $\pm$0.005\\
Ca4227        &    1.049  & $-$0.141 &  $-$0.098 &   $\pm$0.134 &   $\pm$0.069 &   $\pm$0.135 &   $\pm$0.109\\
G4300         &    7.198  & $-$0.181 &   0.126   &   $\pm$0.036 &   $\mp$0.033 &   $\pm$0.106 &   $\pm$0.004\\
Fe4383        &    5.743  & $-$0.272 &   0.029   &   $\pm$0.185 &   $\pm$0.048 &   $\pm$0.243 &   $\pm$0.094\\
Ca4455        &    1.747  & $-$0.063 &   0.004   &   $\pm$0.014 &   $\pm$0.010 &   $\pm$0.128 &   $\pm$0.062\\
Fe4531        &    4.696  & $-$0.064 &   0.054   &   $\mp$0.043 &   $\mp$0.036 &   $\pm$0.308 &   $\pm$0.145\\
C$_{2}$4668   &    2.632  &  0.008   &   0.170   &   $\mp$0.187 &   $\mp$0.168 &   $\pm$0.512 &   $\pm$0.311\\
H$\beta$      &    2.015  &  0.130   &  $-$0.041 &   $\mp$0.078 &   $\mp$0.025 &   $\pm$0.087 &   $\pm$0.009\\
Fe5015        &    7.005  & $-$0.045 &   0.108   &   $\mp$0.141 &   $\mp$0.115 &   $\pm$0.558 &   $\pm$0.272\\
Mg$_{1}$      &    0.103  & $-$0.006 &  $-$0.001 &   $\pm$0.003 &   $\pm$0.002 &   $\pm$0.011 &   $\pm$0.008\\
Mg$_{2}$      &    0.252  & $-$0.019 &  $-$0.010 &   $\pm$0.019 &   $\pm$0.009 &   $\pm$0.021 &   $\pm$0.012\\
Mg$b$         &    3.755  & $-$0.323 &  $-$0.194 &   $\pm$0.482 &   $\pm$0.206 &   $\pm$0.277 &   $\pm$0.145\\
Fe5270        &    3.524  & $-$0.074 &   0.015   &   $\pm$0.041 &   $\mp$0.005 &   $\pm$0.272 &   $\pm$0.122\\
Fe5335        &    4.081  & $-$0.081 &   0.015   &   $\pm$0.045 &   $\mp$0.006 &   $\pm$0.351 &   $\pm$0.192\\
Fe5406        &    2.188  & $-$0.086 &  $-$0.022 &   $\pm$0.052 &   $\pm$0.015 &   $\pm$0.169 &   $\pm$0.103\\
Fe5709        &    1.388  & $-$0.014 &   0.026   &   $\mp$0.031 &   $\mp$0.027 &   $\pm$0.134 &   $\pm$0.075\\
Fe5782        &    0.879  & $-$0.012 &   0.016   &   $\mp$0.013 &   $\mp$0.013 &   $\pm$0.098 &   $\pm$0.061\\
Na D          &    2.813  & $-$0.194 &  $-$0.106 &   $\pm$0.178 &   $\pm$0.071 &   $\pm$0.265 &   $\pm$0.199\\
TiO$_{1}$     &    0.019  & $-$0.005 &  $-$0.004 &   $\pm$0.001 &   $\pm$0.001 &   $\pm$0.002 &   $\pm$0.002\\
TiO$_{2}$     &    0.049  & $-$0.007 &  $-$0.006 &   $\pm$0.000 &   $\pm$0.000 &   $\pm$0.005 &   $\pm$0.004\\
\tableline
\enddata
\end{deluxetable}

%\clearpage

\begin{deluxetable}{lrrrrrrr}
\tablewidth{0pt}
\tablecaption{Results of tests on the 14Gyr SSP. Columns are as for
Table~\ref{tab:4Gtests} \label{tab:14Gtests}}
\tablehead{
\colhead{} & \colhead{} & \multicolumn{2}{c}{$T_{eff}$+100K} & \multicolumn{2}{c}{log$g \pm$0.25} & \multicolumn{2}{c}{[Fe/H]$\pm$0.15}\\
\colhead{} & \colhead{} & \multicolumn{1}{c}{Full iso} & \multicolumn{1}{c}{Giants} & \multicolumn{1}{c}{Full iso} & \multicolumn{1}{c}{Giants} & \multicolumn{1}{c}{Full iso} & \multicolumn{1}{c}{Giants}\\
\colhead{Index} & \multicolumn{1}{c}{Ref} & \multicolumn{1}{c}{(1)}  & \multicolumn{1}{c}{(2)} &
\multicolumn{1}{c}{(3a,b)} & \multicolumn{1}{c}{(4a,b)}  & \multicolumn{1}{c}{(5a,b)} & \multicolumn{1}{c}{(6a,b)}}
\startdata
H$\delta_{F}$ & $-$1.755   &   0.282 &    $-$0.026 &  $\mp$0.083 &   $\pm$0.003 &   $\mp$0.152 &   $\mp$0.102\\
H$\gamma_{F}$ & $-$3.151   &   0.273 &     0.021 &    $\mp$0.132 &   $\mp$0.018 &   $\mp$0.101 &   $\mp$0.093\\
CN$_{1}$      &    0.056   &  $-$0.007 &   0.005 &    $\mp$0.004 &   $\mp$0.001 &   $\pm$0.013 &   $\pm$0.005\\
CN$_{2}$      &    0.114   &  $-$0.010 &   0.003 &    $\mp$0.002 &   $\mp$0.001 &   $\pm$0.017 &   $\pm$0.006\\
Ca4227        &    1.672   &  $-$0.271 &  $-$0.144 &  $\pm$0.234 &   $\pm$0.073 &   $\pm$0.200 &   $\pm$0.134\\
G4300         &    8.287   &  $-$0.035 &   0.069 &    $\mp$0.040 &   $\mp$0.042 &   $\pm$0.015 &   $\pm$0.008\\
Fe4383        &    7.553   &  $-$0.338 &  $-$0.061 &  $\pm$0.198 &   $\pm$0.029 &   $\pm$0.263 &   $\pm$0.095\\
Ca4455        &    2.167   &  $-$0.101 &  $-$0.017 &  $\pm$0.026 &   $\pm$0.008 &   $\pm$0.150 &   $\pm$0.062\\
Fe4531        &    5.319   &  $-$0.098 &   0.025 &    $\mp$0.033 &   $\mp$0.041 &   $\pm$0.322 &   $\pm$0.143\\
C$_{2}$4668   &    3.228   &  $-$0.009 &   0.143 &    $\mp$0.258 &   $\mp$0.168 &   $\pm$0.582 &   $\pm$0.301\\
H$\beta$      &    1.135   &   0.134 &     0.010 &    $\mp$0.072 &   $\mp$0.023 &   $\pm$0.067 &   $\pm$0.008\\
Fe5015        &    7.642   &  $-$0.066 &   0.077 &    $\mp$0.150 &   $\mp$0.116 &   $\pm$0.569 &   $\pm$0.265\\
Mg$_{1}$      &    0.144   &  $-$0.012 &  $-$0.004 &  $\pm$0.006 &   $\pm$0.002 &   $\pm$0.014 &   $\pm$0.007\\
Mg$_{2}$      &    0.338   &  $-$0.029 &  $-$0.015 &  $\pm$0.026 &   $\pm$0.009 &   $\pm$0.025 &   $\pm$0.012\\
Mg$b$         &    4.981   &  $-$0.421 &  $-$0.277 &  $\pm$0.545 &   $\pm$0.192 &   $\pm$0.329 &   $\pm$0.145\\
Fe5270        &    4.107   &  $-$0.112 &  $-$0.014 &  $\pm$0.063 &   $\mp$0.009 &   $\pm$0.273 &   $\pm$0.112\\
Fe5335        &    4.680   &  $-$0.117 &  $-$0.012 &  $\pm$0.052 &   $\mp$0.011 &   $\pm$0.374 &   $\pm$0.174\\
Fe5406        &    2.640   &  $-$0.128 &  $-$0.043 &  $\pm$0.069 &   $\pm$0.012 &   $\pm$0.191 &   $\pm$0.096\\
Fe5709        &    1.557   &  $-$0.021 &   0.020 &    $\mp$0.039 &   $\mp$0.026 &   $\pm$0.139 &   $\pm$0.068\\
Fe5782        &    1.019   &  $-$0.021 &   0.012 &    $\mp$0.016 &   $\mp$0.014 &   $\pm$0.107 &   $\pm$0.055\\
Na D          &    3.711   &  $-$0.306 &  $-$0.160 &  $\pm$0.269 &   $\pm$0.067 &   $\pm$0.317 &   $\pm$0.190\\
TiO$_{1}$     &    0.022   &  $-$0.006 &  $-$0.004 &  $\pm$0.001 &   $\pm$0.001 &   $\pm$0.002 &   $\pm$0.002\\
TiO$_{2}$     &    0.056   &  $-$0.009 &  $-$0.006 &  $\pm$0.000 &   $\mp$0.000 &   $\pm$0.006 &   $\pm$0.004\\
\tableline
\enddata
\end{deluxetable}

%\clearpage 

\begin{deluxetable}{lccc}
\tablewidth{0pt}
\tablecaption{Mean observational errors on line-strengths from 3 sources: \citet{sanbla06} (SB06), 
\citet{sanbla09} (SB09) and \citet{trager08} (T08) \label{obserrors}}
\tablehead{
\colhead{Index} & \colhead{SB06}  & \colhead{SB09} & \colhead{T08}}
\startdata
H$\delta_{F}$ & 0.107 & 0.303 & 0.100 \\
H$\gamma_{F}$ & 0.108 & 0.314 & 0.111 \\
CN$_{1}$      & 0.006 &  ---  & 0.004 \\
CN$_{2}$      & 0.007 & 0.016 & 0.004 \\
Ca4227        & 0.076 & 0.250 & 0.067 \\
G4300         & 0.184 & 0.438 & 0.128 \\
Fe4383        & 0.208 & 0.721 & 0.152 \\
Ca4455        & 0.093 & 0.396 & 0.086 \\
Fe4531        & 0.131 & 0.613 & 0.107 \\
C$_{2}$4668   & 0.253 &  ---  & 0.165 \\
H$\beta$      & 0.079 &  ---  & 0.066 \\
Fe5015        & 0.189 &  ---  & 0.142 \\
Mg$b$         & 0.143 &  ---  & 0.068 \\
Fe5270        &  ---  &  ---  & 0.076 \\
Fe5335        &  ---  &  ---  & 0.085 \\
\tableline
\enddata
\end{deluxetable}

\bibliographystyle{apj}
\bibliography{refs}

\begin{thebibliography}{37}
\expandafter\ifx\csname natexlab\endcsname\relax\def\natexlab#1{#1}\fi

\bibitem[{{Alonso} {et~al.}(1996{\natexlab{a}}){Alonso}, {Arribas}, \&
  {Martinez-Roger}}]{alonso96b}
{Alonso}, A., {Arribas}, S., \& {Martinez-Roger}, C. 1996{\natexlab{a}}, \aaps,
  117, 227

\bibitem[{{Alonso} {et~al.}(1996{\natexlab{b}}){Alonso}, {Arribas}, \&
  {Martinez-Roger}}]{alonso96}
---. 1996{\natexlab{b}}, \aap, 313, 873

\bibitem[{{Alonso} {et~al.}(1999{\natexlab{a}}){Alonso}, {Arribas}, \&
  {Mart{\'{\i}}nez-Roger}}]{alonso99b}
{Alonso}, A., {Arribas}, S., \& {Mart{\'{\i}}nez-Roger}, C. 1999{\natexlab{a}},
  \aaps, 139, 335

\bibitem[{{Alonso} {et~al.}(1999{\natexlab{b}}){Alonso}, {Arribas}, \&
  {Mart{\'{\i}}nez-Roger}}]{alonso99}
---. 1999{\natexlab{b}}, \aaps, 140, 261

\bibitem[{{Bruzual}(2007)}]{bruzual}
{Bruzual}, A.~G. 2007, in IAU Symposium, Vol. 241, IAU Symposium, ed.
  A.~{Vazdekis} \& R.~F. {Peletier}, 125--132

\bibitem[{{Bruzual} \& {Charlot}(2003)}]{bc03}
{Bruzual}, G. \& {Charlot}, S. 2003, \mnras, 344, 1000

\bibitem[{{Carter} {et~al.}(2009){Carter}, {Smith}, {Percival}, {Baldry},
  {Collins}, {James}, {Salaris}, {Simpson}, {Stott}, \& {Mobasher}}]{carter}
{Carter}, D., {Smith}, D.~J.~B., {Percival}, S.~M., {Baldry}, I.~K., {Collins},
  C.~A., {James}, P.~A., {Salaris}, M., {Simpson}, C., {Stott}, J.~P., \&
  {Mobasher}, B. 2009, \mnras, 397, 695

\bibitem[{{Casagrande} {et~al.}(2006){Casagrande}, {Portinari}, \&
  {Flynn}}]{casagrande}
{Casagrande}, L., {Portinari}, L., \& {Flynn}, C. 2006, \mnras, 373, 13

\bibitem[{{Cenarro} {et~al.}(2008){Cenarro}, {Cervantes}, {Beasley},
  {Mar{\'{\i}}n-Franch}, \& {Vazdekis}}]{cenarro}
{Cenarro}, A.~J., {Cervantes}, J.~L., {Beasley}, M.~A., {Mar{\'{\i}}n-Franch},
  A., \& {Vazdekis}, A. 2008, \apjl, 689, L29

\bibitem[{{Coelho} {et~al.}(2007){Coelho}, {Bruzual}, {Charlot}, {Weiss},
  {Barbuy}, \& {Ferguson}}]{coelho07}
{Coelho}, P., {Bruzual}, G., {Charlot}, S., {Weiss}, A., {Barbuy}, B., \&
  {Ferguson}, J.~W. 2007, \mnras, 382, 498

\bibitem[{{Girardi} {et~al.}(2000){Girardi}, {Bressan}, {Bertelli}, \&
  {Chiosi}}]{girardi2000}
{Girardi}, L., {Bressan}, A., {Bertelli}, G., \& {Chiosi}, C. 2000, \aaps, 141,
  371

\bibitem[{{Gonz{\'a}lez Hern{\'a}ndez} \& {Bonifacio}(2009)}]{gonz}
{Gonz{\'a}lez Hern{\'a}ndez}, J.~I. \& {Bonifacio}, P. 2009, \aap, 497, 497

\bibitem[{{James} {et~al.}(2006){James}, {Salaris}, {Davies}, {Phillipps}, \&
  {Cassisi}}]{paj}
{James}, P.~A., {Salaris}, M., {Davies}, J.~I., {Phillipps}, S., \& {Cassisi},
  S. 2006, \mnras, 367, 339

\bibitem[{{Kotulla} {et~al.}(2009){Kotulla}, {Fritze}, {Weilbacher}, \&
  {Anders}}]{galev09}
{Kotulla}, R., {Fritze}, U., {Weilbacher}, P., \& {Anders}, P. 2009, \mnras,
  396, 462

\bibitem[{{Maraston}(2005)}]{maraston05}
{Maraston}, C. 2005, \mnras, 362, 799

\bibitem[{{Martins} \& {Coelho}(2007)}]{martco}
{Martins}, L.~P. \& {Coelho}, P. 2007, \mnras, 381, 1329

\bibitem[{{Munari} {et~al.}(2005){Munari}, {Sordo}, {Castelli}, \&
  {Zwitter}}]{munari}
{Munari}, U., {Sordo}, R., {Castelli}, F., \& {Zwitter}, T. 2005, \aap, 442,
  1127

\bibitem[{{Percival} {et~al.}(2009){Percival}, {Salaris}, {Cassisi}, \&
  {Pietrinferni}}]{basti4}
{Percival}, S.~M., {Salaris}, M., {Cassisi}, S., \& {Pietrinferni}, A. 2009,
  \apj, 690, 427

\bibitem[{{Pietrinferni} {et~al.}(2004){Pietrinferni}, {Cassisi}, {Salaris}, \&
  {Castelli}}]{basti1}
{Pietrinferni}, A., {Cassisi}, S., {Salaris}, M., \& {Castelli}, F. 2004, \apj,
  612, 168

\bibitem[{{Ram{\'{\i}}rez} \& {Mel{\'e}ndez}(2005)}]{ramirez}
{Ram{\'{\i}}rez}, I. \& {Mel{\'e}ndez}, J. 2005, \apj, 626, 446

\bibitem[{{Salaris} \& {Cassisi}(2007)}]{salcas}
{Salaris}, M. \& {Cassisi}, S. 2007, \aap, 461, 493

\bibitem[{{S{\'a}nchez-Bl{\'a}zquez}
  {et~al.}(2006{\natexlab{a}}){S{\'a}nchez-Bl{\'a}zquez}, {Gorgas}, {Cardiel},
  \& {Gonz{\'a}lez}}]{sanbla06}
{S{\'a}nchez-Bl{\'a}zquez}, P., {Gorgas}, J., {Cardiel}, N., \& {Gonz{\'a}lez},
  J.~J. 2006{\natexlab{a}}, \aap, 457, 787

\bibitem[{{S{\'a}nchez-Bl{\'a}zquez} {et~al.}(2009){S{\'a}nchez-Bl{\'a}zquez},
  {Jablonka}, {Noll}, {Poggianti}, {Moustakas}, {Milvang-Jensen}, {Halliday},
  {Arag{\'o}n-Salamanca}, {Saglia}, {Desai}, {De Lucia}, {Clowe}, {Pell{\'o}},
  {Rudnick}, {Simard}, {White}, \& {Zaritsky}}]{sanbla09}
{S{\'a}nchez-Bl{\'a}zquez}, P., {Jablonka}, P., {Noll}, S., {Poggianti}, B.~M.,
  {Moustakas}, J., {Milvang-Jensen}, B., {Halliday}, C.,
  {Arag{\'o}n-Salamanca}, A., {Saglia}, R.~P., {Desai}, V., {De Lucia}, G.,
  {Clowe}, D.~I., {Pell{\'o}}, R., {Rudnick}, G., {Simard}, L., {White},
  S.~D.~M., \& {Zaritsky}, D. 2009, \aap, 499, 47

\bibitem[{{S{\'a}nchez-Bl{\'a}zquez}
  {et~al.}(2006{\natexlab{b}}){S{\'a}nchez-Bl{\'a}zquez}, {Peletier},
  {Jim{\'e}nez-Vicente}, {Cardiel}, {Cenarro}, {Falc{\'o}n-Barroso}, {Gorgas},
  {Selam}, \& {Vazdekis}}]{miles1}
{S{\'a}nchez-Bl{\'a}zquez}, P., {Peletier}, R.~F., {Jim{\'e}nez-Vicente}, J.,
  {Cardiel}, N., {Cenarro}, A.~J., {Falc{\'o}n-Barroso}, J., {Gorgas}, J.,
  {Selam}, S., \& {Vazdekis}, A. 2006{\natexlab{b}}, \mnras, 371, 703

\bibitem[{{Schiavon}(2007)}]{schi07}
{Schiavon}, R.~P. 2007, \apjs, 171, 146

\bibitem[{{Schiavon} {et~al.}(2002){Schiavon}, {Faber}, {Rose}, \&
  {Castilho}}]{schi02}
{Schiavon}, R.~P., {Faber}, S.~M., {Rose}, J.~A., \& {Castilho}, B.~V. 2002,
  \apj, 580, 873

\bibitem[{{Schiavon} {et~al.}(2004){Schiavon}, {Rose}, {Courteau}, \&
  {MacArthur}}]{schi04}
{Schiavon}, R.~P., {Rose}, J.~A., {Courteau}, S., \& {MacArthur}, L.~A. 2004,
  \apjl, 608, L33

\bibitem[{{Schiavon} {et~al.}(2005){Schiavon}, {Rose}, {Courteau}, \&
  {MacArthur}}]{schi05}
---. 2005, \apjs, 160, 163

\bibitem[{{Soubiran} {et~al.}(1998){Soubiran}, {Katz}, \& {Cayrel}}]{skc}
{Soubiran}, C., {Katz}, D., \& {Cayrel}, R. 1998, \aaps, 133, 221

\bibitem[{{Thomas} {et~al.}(2003){Thomas}, {Maraston}, \& {Bender}}]{tmb03}
{Thomas}, D., {Maraston}, C., \& {Bender}, R. 2003, \mnras, 339, 897

\bibitem[{{Trager} {et~al.}(2008){Trager}, {Faber}, \& {Dressler}}]{trager08}
{Trager}, S.~C., {Faber}, S.~M., \& {Dressler}, A. 2008, \mnras, 386, 715

\bibitem[{{Trager} {et~al.}(2000){Trager}, {Faber}, {Worthey}, \&
  {Gonz{\'a}lez}}]{trager2000}
{Trager}, S.~C., {Faber}, S.~M., {Worthey}, G., \& {Gonz{\'a}lez}, J.~J. 2000,
  \aj, 119, 1645

\bibitem[{{Trager} {et~al.}(1998){Trager}, {Worthey}, {Faber}, {Burstein}, \&
  {Gonzalez}}]{trager98}
{Trager}, S.~C., {Worthey}, G., {Faber}, S.~M., {Burstein}, D., \& {Gonzalez},
  J.~J. 1998, \apjs, 116, 1

\bibitem[{{Vazdekis} {et~al.}(2001){Vazdekis}, {Salaris}, {Arimoto}, \&
  {Rose}}]{vaz01}
{Vazdekis}, A., {Salaris}, M., {Arimoto}, N., \& {Rose}, J.~A. 2001, \apj, 549,
  274

\bibitem[{{V{\'a}zquez} \& {Leitherer}(2005)}]{starb99}
{V{\'a}zquez}, G.~A. \& {Leitherer}, C. 2005, \apj, 621, 695

\bibitem[{{Worthey}(1994)}]{worthey94}
{Worthey}, G. 1994, \apjs, 95, 107

\bibitem[{{Worthey} \& {Ottaviani}(1997)}]{worthott}
{Worthey}, G. \& {Ottaviani}, D.~L. 1997, \apjs, 111, 377

\end{thebibliography}

\end{document}